\documentclass[twocolumn,floatfix,showpacs,amsmath,amssymb,nofootinbib,superscriptaddress,
prd]
{revtex4}
\pdfoutput=1

\usepackage{graphicx}
\usepackage{epsfig}
\usepackage{bm}
\usepackage{amssymb}
\usepackage{float}
\usepackage{amsmath}
\usepackage{dcolumn}
\usepackage{cancel}
\usepackage[colorlinks]{hyperref}
\usepackage[usenames,dvipsnames]{color}

\newcommand{\HCd}{\mathcal{H}}

\def\HCdt0{\tilde{\HCd}_{0}}

\newcommand{\afffias}{Frankfurt Institute for Advanced Studies (FIAS), 
Ruth-Moufang-Strasse~1, 60438 Frankfurt am Main, Germany}
\newcommand{\affbgu}{Physics Department, Ben-Gurion University of the Negev, Beer-Sheva 
84105, Israel}
\newcommand{\affbahamas}{Bahamas Advanced Study Institute and Conferences, 4A Ocean 
Heights, Hill View Circle, Stella Maris, Long Island, The Bahamas}
\newcommand{\affgsi}{GSI Helmholtzzentrum f\"ur Schwerionenforschung GmbH, 
Planckstrasse~1, 64291 Darmstadt, Germany}
\newcommand{\affgreece}{Department of Physics, National Technical University of Athens, 
Zografou Campus GR 157 73, Athens, Greece}
\newcommand{\affchaina}{Department of Astronomy, School of Physical Sciences, University 
of Science
and Technology of China, Hefei 230026, P.R. China}

\newcommand{\affchainaa}{Chongqing University of Posts \& Telecommunications, Chongqing, 
400065, P.R.
China}

\newcommand{\affguni}{Fachbereich Physik, Goethe-Universitat, Max-von-Laue-Strasse 1, 
60438 Frankfurt am Main, Germany}

\begin{document}

\title{ Inflation from fermions with curvature-dependent mass}
 
\author{D. Benisty}
\email{benidav@post.bgu.ac.il}
\affiliation{\afffias}\affiliation{\affbgu}
\author{E.I. Guendelman}
\email{guendel@bgu.ac.il}
\affiliation{\affbgu}\affiliation{\afffias}\affiliation{\affbahamas}
\author{E. N. Saridakis}
\email{msaridak@phys.uoa.gr}
\affiliation{\affgreece}\affiliation{\affchaina}\affiliation{\affchainaa}
\author{H.~Stoecker}
\email{stoecker@fias.uni-frankfurt.de}
\affiliation{\afffias}\affiliation{\affguni}\affiliation{\affgsi}
\author{J.~Struckmeier}
\email{struckmeier@fias.uni-frankfurt.de}
\affiliation{\afffias}\affiliation{\affguni}
\author{D.~Vasak}
\email{vasak@fias.uni-frankfurt.de}
\affiliation{\afffias}


\begin{abstract}
A model of inflation realization driven by fermions with curvature-dependent mass is 
studied.
Such a term is derived from the Covariant Canonical Gauge Theory of gravity 
(CCGG) incorporating Dirac fermions.
We obtain an initial de~Sitter phase followed by a successful exit, and moreover we 
acquire the subsequent thermal history, with an effective matter era, followed finally by 
a dark-energy epoch.
This behavior is a result of the effective ``weakening'' of gravity at early times, due to 
the increased curvature-dependent fermion mass.
Investigating the scenario at the perturbation level, using the correct coupling 
parameter, the scalar spectral index and tensor-to-scalar ratio are obtained in agreement 
with Planck observations.
Moreover the BBN constraints are satisfied too.
The efficiency of inflation from fermions with curvature-dependent mass, at both the 
background and perturbation level, reveals the capabilities of the scenario and makes it a 
good candidate for the description of nature. 
\end{abstract}

\maketitle

\section{Introduction}

The developments in cosmology have been influenced to a great extent by the idea of 
inflation 
\cite{Starobinsky:1979ty,Kazanas:1980tx,Starobinsky:1980te,Guth:1980zm,Linde:1981mu,
Albrecht:1982wi,Blau:1986cw}, which provides an attractive 
scenario for the solution of the fundamental puzzles of the old Big Bang paradigm, such as 
the horizon and the flatness problems. Additionally, inflation was proved crucial in 
providing a framework for the generation of primordial density perturbations 
\cite{Mukhanov:1981xt,Guth:1982ec}.
Although the inflationary scenario is very attractive, it has been recognized that a 
successful implementation requires special restrictions on the underlying dynamics.
The inflationary mechanism can be achieved in several different ways, considering 
primordial scalar fields \cite{Lidsey:1995np,Bassett:2005xm} or geometric corrections into 
the effective gravitational action \cite{Nojiri:2003ft}.

On the other hand, it is known that Hamiltonian formulations of a theory may have 
theoretical  advantages. One such framework is  the covariant canoninal gauge theory of 
gravity (CCGG) \cite{Struckmeier:2017vkf}.
The CCGG ensures by construction that the action principle is maintained in its form by 
requiring that all transformations of a given system are canonical.
The imposed requirement of invariance of the original action integral with respect to 
local transformations in curved spacetime is achieved by introducing additional degrees of 
freedom, namely the gauge fields \cite{Struckmeier:2017vkf}.
In these lines, in \cite{Benisty:2018ufz,Benisty:2018efx,Benisty:2018fgu,Vasak:2018lhz} 
quadratic Riemann theories by the covariant Hamiltonian approach were formulated, which 
were shown to lead to inflationary models 
\cite{Myrzakulov:2014hca} based on the correspondence between the metric affine 
(Palatini) formalism and the metric formalism \cite{Benisty:2018fgu,Benisty:2018efx}.
 
The covariant Hamiltonian formulation was recently implemented to include Dirac 
fermions \cite{Struckmeier:2018psp}.
This covariant Hamiltonian incorporates additional terms with a new dependence for 
fermions, namely the effective mass depends linearly on the curvature through
\begin{equation}\label{mass}
m_{\text{eff}} = m_0 + \frac{\xi}{6\kappa^2} \mathcal{R},
\end{equation}
where $m_0$ is the usual fermion rest mass, $\xi$ is the coupling to the Ricci scalar 
$\mathcal{R}$ with dimensions $[L]^3$, and $\kappa^2 = 8\pi G_N$. 
For $\xi \rightarrow 0$, the covariant Hamiltonian  approach for fermions reduces to that 
of standard Dirac fermions. 

The curvature-dependence mass of the fermions is expected to be significant in regimes 
where the curvature is significant, such as close to black holes 
\cite{Struckmeier:2018psp} or new description for Neutrino \cite{Onofrio:2013iya}. 
However, one can deduce that this curvature-dependent correction can be important also in 
the early universe, where it is known that the Ricci scalar acquires large values.
Hence, it would be interesting to investigate the effect of the fermion 
curvature-dependent mass in the early universe.
In particular, we desire to examine whether it can drive a successful inflation. Indeed, 
as we will see, such a novel coupling can both drive a successful inflation at the 
background level, accompanied by a successful exit and the 
subsequent thermal history of the universe, but it can also be very efficient  at the 
perturbation level, giving rise to a scalar spectral index and a tensor-to-scalar ratio in 
agreement with observations.

The plan of the work is the following: In Section \ref{themodel} we present fermionic 
cosmology with curvature-dependent mass, extracting the equations of motion.
In Section~\ref{Backgroundevolution} we apply the scenario in the early universe, 
obtaining the inflation realization at the background level.
In Section~\ref{Perturbations} we examine the perturbation-related observables such as the 
spectral index and the tensor-to-scalar ratio.
Finally, in Section \ref{Conclusions} we summarize our results.

\section{The model}
\label{themodel}

In this section we construct the scenario of fermionic cosmology with curvature-dependent 
mass.
We first briefly review the basics of the  covariant Hamiltonian incorporation of fermions 
in curved spacetime, and then we present the Lagrangian of the model, extracting the 
equations of motion.

\subsection{Spinors in curved spacetime}

Fermionic sources in general relativity  were studied in detail in 
\cite{Ellis:1999sf,Ribas:2005vr,Samojeden:2010rs}, especially for cosmological 
applications~\cite{Myrzakulov:2010du,Chimento:2007fx,Ribas:2016ulz,Ribas:2007qm,
Hossain:2014zma,Basilakos:2015yoa,Geng:2017mic,Paganini:2018hof}.
The tetrad formalism was used to combine the gauge group of general relativity with a 
spinor matter field.
The tetrad $e^{a}_{\mu}$ and the metric $g_{\mu\nu} $ tensors are related through
\begin{equation}
g_{\mu\nu} = e^{a}_{\mu} e^{b}_{\nu} \eta_{a b}, \quad a,b = 0,1,2,3,
\label{metrictetrad}
\end{equation} 
with Latin indices referring to the local inertial frame endowed with the Minkowski metric 
$\eta_{ab}$, while Greek indices denote the (holonomic) basis of the manifold.

The  spinor field Lagrangian in curved torsion-free space-time reads as
\begin{equation}
\mathcal{L}_f=\frac{i}{2}\left[ \bar{\psi}\,\Gamma^\mu
D_\mu\psi-\left(D_\mu\bar{\psi}\right)\Gamma^\mu\psi\right]-m_{\text{eff}} \, 
\bar{\psi}\psi,
\end{equation}
where $\bar{\psi} = \psi^{+} \gamma^0$ is the adjoint spinor field and $m$ is the 
fermionic rest mass, as defined in~(\ref{mass}).
In curved space time the Dirac matrices 
are replaced by their generalized counterparts $\Gamma^{\mu} = e^{\mu}_{a} \gamma^{a}$, 
which satisfy the extended Clifford algebra 
$\frac{1}{2}\left(\Gamma^{\mu}\Gamma^{\nu}+\Gamma^{\nu}\Gamma^{\mu} \right)= g^{\mu\nu}$.
Thus, the ordinary derivatives are replaced by their covariant versions
\begin{equation}
D_\mu\psi= \partial_\mu\psi-\Omega_\mu\psi.
\end{equation}
Furthermore, the metric compatibility condition implies that the spin connection 
$\Omega_\mu$ is given by
\begin{equation}\label{Omega}
\Omega_\mu=\frac{1}{4}g_{\beta \nu} \left[\bar\Gamma^{\nu}_{\alpha\mu} - e^{\nu}_{j} 
\partial_{\mu} e^{j}_{\alpha} \right] \Gamma^{\beta} \Gamma^{\alpha},
\end{equation}
with ${\bar\Gamma}^\nu_{\sigma\lambda}$  the Christoffel symbols. 
The original CCGG formulation assumes that the affine and spin connections are fields 
independent of the metric (i.e.\ the metric-affine or the Palatini formulation) 
\cite{Struckmeier:2018psp}, however in this work we use the metric compatibility derived 
from the action, that allows for the substitution (\ref{Omega}).

The action of fermions in the gravitational background of general relativity is then 
\begin{equation}
\mathcal{S} = \int d^4x \sqrt{-g} \left[  \frac{1}{2 \kappa^2} \left(\mathcal{R} - 
2\Lambda 
\right)+\mathcal{L}_{f} \right],
\label{totaction}
\end{equation}
where $\mathcal{R}$ is the Ricci scalar and $\Lambda$ the  cosmological constant. 
Note that concerning the fermions we have assumed a minimal coupling with gravity, while 
their inertial mass $m_{\text{eff}}$ is given by (\ref{mass}). 
Here, rather then applying the complete CCGG action, we simplify the analysis is order to 
illuminate the impact of the effective spinor mass in the conventional Einstein-Hilbert 
theory in the metric formulation.  

\subsection{Field equations}

Let us now extract the field equations of the action (\ref{totaction}).
Variation with respect to the spinor field yields the generalized Dirac 
equations:
\begin{equation}\label{dirac}
\begin{split}
i \Gamma^{\mu} D_{\mu}\psi - m_\text{eff} \psi = 0 \\
i D_{\mu}\bar{\psi} \Gamma^{\mu} + m_\text{eff} \bar{\psi} = 0.
\end{split}
\end{equation}
Moreover,  variation with respect to the metric leads to the field equations
\begin{equation}
\label{Eom0}
\frac{1}{\kappa^2}G^{\mu\nu} = T^{\mu\nu}_{(f)}+ \frac{\xi}{3\kappa^2} \left[G^{\mu\nu} 
\bar{\psi} \psi +  \Box^{\mu\nu} (\bar{\psi} \psi)\right] + g^{\mu\nu} 
\frac{\Lambda}{\kappa^2},
\end{equation}
where $G^{\mu\nu}$ is the Einstein tensor, and 
$
\Box_{\mu\nu} = g_{\mu\nu} \Box-\nabla_\mu\nabla_\nu$. In the above equations 
$T^{\mu\nu}_{(f)}$ is the kinetic part of the spinor fields, given by
\begin{eqnarray}
&&\!\!\!\!\!\!\!\!\!\!\!\!\!\!\!\!
T^{\mu\nu}_{(f)} = \frac{i}{4} \left[ \bar{\psi} \Gamma^{(\mu} D^{\nu)}\psi - 
D^{(\nu}\bar{\psi}\Gamma^{\mu)} \psi \right]\nonumber \\
&&\
- g^{\mu\nu} \left\{ \frac{i}{2} 
\left[\bar{\psi}\Gamma^\lambda D_{\lambda} \psi -D_{\lambda} \bar{\psi}\Gamma^\lambda  
\psi \right] - m_{\text{eff}} \,  \bar{\psi} \psi \right\}.
\end{eqnarray}
Equation (\ref{Eom0}) can be re-written as 
\begin{eqnarray}\label{metVar}
 &&
 \!\!\!\!\!\!\! \!\!\!\!\!\!\!
 \frac{1}{\kappa^2} \left(1-\frac{\xi}{3} \, \bar\psi \psi \right)G^{\mu\nu} = 
\frac{i}{4} 
\left[ \bar{\psi} \Gamma^{(\mu} D^{\nu)}\psi - D^{(\nu}\bar{\psi}\Gamma^{\mu)} \psi 
\right]\nonumber \\
&&\ \ \ \ \ 
- g^{\mu\nu} \left\{ \frac{i}{2} \left[\bar{\psi}\Gamma^\lambda D_{\lambda} 
\psi 
-D_{\lambda} \bar{\psi}\Gamma^\lambda  \psi \right] - m_{\text{eff}} \, \bar{\psi} \psi 
\right\} \nonumber\\ 
&&\ \ \ \ \     - \frac{\xi}{3 \kappa^2} \, \Box^{\mu\nu} (\bar{\psi} \psi) + g^{\mu\nu} 
\frac{\Lambda}{\kappa^2}.
\end{eqnarray}
Finally, note 
that the phase invariance of the spinor field
\begin{equation}
\psi \rightarrow e^{i\theta} \psi, \quad \bar\psi \rightarrow e^{-i\theta} \bar\psi,
\end{equation}
through the Noether's theorem leads to the conserved current
\begin{equation}\label{current}
 j^{\mu}_{;\mu}=0,
\end{equation}
where $j^{\mu} = \bar{\psi} \gamma^\mu \psi $, which proves convenient in simplifying our 
analysis.

\section{Background evolution}
\label{Backgroundevolution}

In this section we apply the above fermionic model at a cosmological framework, focusing 
on early-time universe, and in particular on the inflationary realization. As usual 
we neglect standard matter, i.e. we incorporate only the fermions with 
curvature-dependent mass. We consider 
the  Friedman-Lemaitre-Robertson-Walker (FLRW) homogeneous and 
isotropic  metric 
\begin{equation}\label{metric}
ds^2 =-dt^2 +a(t)^2 \left(dx^2+dy^2+dz^2\right),
\end{equation}
with the scale factor $a(t)$, and thus through (\ref{metrictetrad}) the tetrad
components are found to be
\begin{equation}
e^{\mu}_{0} = \delta^{\mu}_{0}, \quad  e^{\mu}_{i} = \frac{1}{a(t)}\delta^{\mu}_{i}.
\end{equation}
Moreover, the covariant version of the Dirac matrices are
\begin{equation}
\Gamma^{0} = \gamma^{0}, \quad \Gamma^{i} =\frac{1}{a(t)}\gamma^{i},
\end{equation}
while the spin connection becomes
\begin{equation}
\Omega_{0} = \gamma^{0}, \quad \Omega_i = \frac{1}{2} \dot{a}(t) \gamma^{i}\gamma^0.
\end{equation}
The Dirac equation (\ref{dirac}) then 
becomes
\begin{equation}
\dot{\psi }+ \frac{3}{2} H \psi + i m_{\text{eff}} \, \gamma^0 \psi = 0,
\label{gendirac1}
\end{equation}
and similarly for  $\bar{\psi}$, with
\begin{equation}
m_{\text{eff}} = m_0 + 
\frac{\xi}{\kappa^2}(\dot{H}+2H^2) ,
\label{meffrel}
\end{equation}
and $H=\dot{a}/a$ the Hubble function.
For an isotropic and homogeneous universe the spinor field is exclusively a function of 
time, namely
\begin{equation}
\psi = \begin{pmatrix}
    \psi_0(t) \,e^{i \phi(t)} \\
    0 \\
    0 \\
    0
\end{pmatrix},
\end{equation}
with $\phi(t)$   the phase of the state, where we have considered the 
fermions to be at the minimal state.  Introducing this ansatz into the 
generalized Dirac equation (\ref{gendirac1}) yields the relation between the phase of the 
fermion and the effective mass, namely
\begin{equation}
\frac{d}{dt}\phi(t) = - m_\text{eff}.
\end{equation}

An elegant way to extract the solution for the spinor field is by making use of the 
conserved current 
(\ref{current}). For the FLRW metric the conserved current becomes
\begin{equation}
\frac{1}{\sqrt{-g}}\partial_\mu \left(\sqrt{-g} j^\mu \right) = 0 \quad \Rightarrow \quad 
\frac{1}{a^3}\partial_t \left(a^3 j^0 \right) = 0,
\end{equation}
which then leads to a dust-like 
behavior for the absolute value of the spinor field, i.e.
\begin{equation}
\bar\psi \psi = \psi_0 (t)^2 = \frac{C}{a^3},
\label{psipsibar}
\end{equation}
with the particle number density $C > 0$. The quantity $m_0 C$ is thus the total gas 
energy density.

Let us now turn to the field equations (\ref{metVar}). Inserting the FLRW metric 
(\ref{metric}) they give rise to the Friedmann equations 
\begin{equation}\label{curFin}
- H^2\xi \psi_0^2+3 H^2-2 H \xi  \psi_0 \dot{\psi}_0 -\Lambda -\kappa^2 m_0 \psi_0^2= 0,
\end{equation}
\begin{eqnarray}
&&
\!\!\!\!\!\!\!\!\!\!\!\!\!\!\!\!\! 
\left(3-\frac{2 C \xi }{a^3}\right)H^2 +\left(\frac{C \xi }{3 a^3}+2\right) 
\dot{H}\nonumber\\
&&
\ \ \ \ \ \ \ \ \ \ \ \ \,
+\frac{C\kappa ^2}{2 a^3} \left(H-2 m_0+2 \phi \dot{\phi}\right)-\Lambda  = 
0.
 \end{eqnarray}
 Combining (\ref{psipsibar}) with 
(\ref{curFin}) we obtain
the convenient form
\begin{equation}\label{Friedmann}
\frac{3H^2}{\kappa^2}+\frac{2\tilde{\xi}H^2}{\kappa^2a^3}
= \frac{\rho_{f0}}{a^3}+\Lambda,
\end{equation}
with the definitions $\rho_{f0}= m_0 C$ and $\tilde{\xi} = C \xi$.
Finally, since $\kappa^2 = 8\pi G_N$,  from Eq.  (\ref{Friedmann})  we   deduce 
that the curvature-dependent fermion mass term  induces an effective Newtonian 
constant
\begin{equation}
\label{Geff}
\frac{G_\text{eff}}{G_N} = \frac{3 a^3}{3 a^3 + 2\tilde{\xi}}.
\end{equation}

We proceed by investigating the inflation realization in the above construction. From 
now on we set  $\kappa^2 = 8\pi G_N=1$. For small scale factors, such that
\begin{equation}
a^3\ll \frac{\rho_{f0}}{\Lambda} = \frac{m_0 C}{\Lambda}\equiv
s, \quad a^3\ll \tilde{\xi},
\label{approximation}
\end{equation}
the modified Friedmann equation (\ref{Friedmann}) accepts the de Sitter solution 
 $a \sim e^{H_{inf} t}$, with 
\begin{equation}\label{Hinf}
H_{inf}  = \sqrt{\frac{\rho_{f0}}{2\tilde{\xi}}}.
\end{equation}
Thus, we observe that the inflation phase is driven by the effectively dust-like 
fermionic component. 
However, after a suitable inflationary expansion the scale factor increases significantly 
and the approximation (\ref{approximation}) breaks down, leading to a decrease of the 
Hubble function and thus to a successful inflationary exit. At later times the 
dust-like terms of the Friedmann equation  (\ref{Friedmann}) will 
dominate, driving the matter epoch. Finally, at late-times, the cosmological constant 
term will dominate in 
(\ref{Friedmann}) and the Hubble constant asymptotically approaches the value
\begin{equation}
H_{\infty} = \sqrt{\frac{\Lambda}{3}}.
\end{equation}
 This thermal history of the universe is in 
agreement with observations. However, we should mention here that a scenario fully 
consistent with the observed universe evolution and complete, should include a   
description of the post-inflationary phase of reheating, that could produce the radiation 
sector, which could then dominate and thus drive the necessary radiation era before the 
matter one. The detailed elaboration of this intermediate evolution lies beyond the scope 
of the present work and will be investigated in a separate project.

We now present a convenient way to extract the above results. 
In particular, we introduce an  ``effective potential''  through the relation $\dot{a}^2 
+ V(a) = 0$ (i.e. the 
``kinetic energy'' 
$\dot{a}^2$ and the potential energy $V(a)$ add up to zero), which 
using the Friedmann equation  (\ref{Friedmann}) leads to \cite{effectivePot}
\begin{equation}
\label{potential}
 V(a) =  -a^2 \Lambda\left(\frac{s+a^3}{2 \tilde{\xi} + 3 a^3}\right).
\end{equation}
 \begin{figure}[ht]
 	\centering
\includegraphics[width=0.49\textwidth]{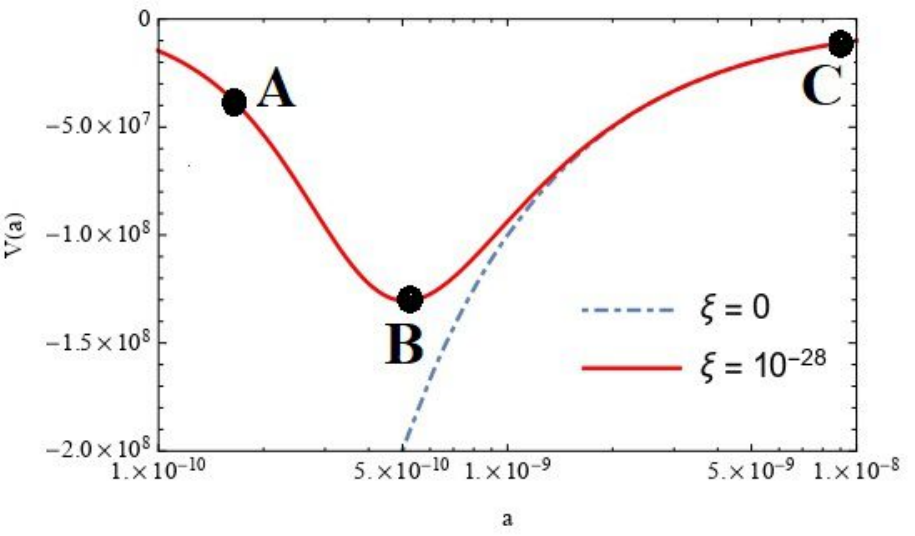}
\caption{{\it{The ``effective potential'' from Eq.~(\ref{potential})}}}
\label{fig1}
\end{figure}
The effective potential is presented in Fig.~\ref{fig1}. 
The initial point $V(a=0) = 0$ incorporates  time scales beyond Planck time, and thus 
our discussion starts at point $A$, which corresponds to the inflationary initial scale 
factor $a_i$ (which is calculated in the next section). The potential contains one 
minimal point $B$ 
and one maximal point $C$, with the scale factors
\begin{equation}
a_{B,C}^3 = \frac{1}{12}\left(3 s-10 \tilde{\xi} \pm \sqrt{3 s-50\tilde{\xi} } \sqrt{3 
s-2 
\tilde{\xi}  }\right),
\end{equation}
where the $+$ sign corresponds to point $C$. One can easily see that point $B$ is the 
point where inflation ends, and the universe 
enters the effective dust epoch. This phase holds up to point $C$, 
after which the universe enters into the late-time, dark energy regime.

\begin{figure}[ht]
	\centering
\includegraphics[width=0.475\textwidth]{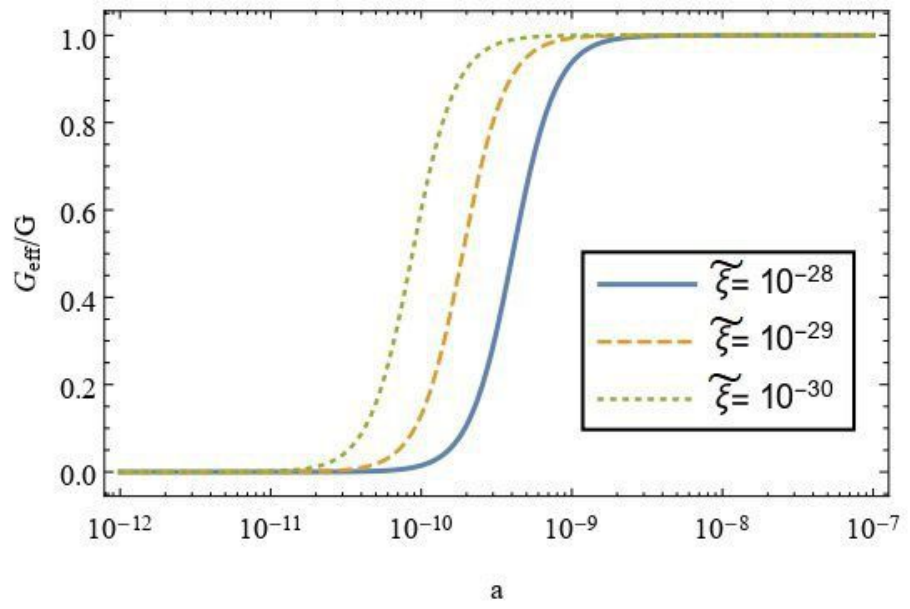}
\\\includegraphics[width=0.475\textwidth]{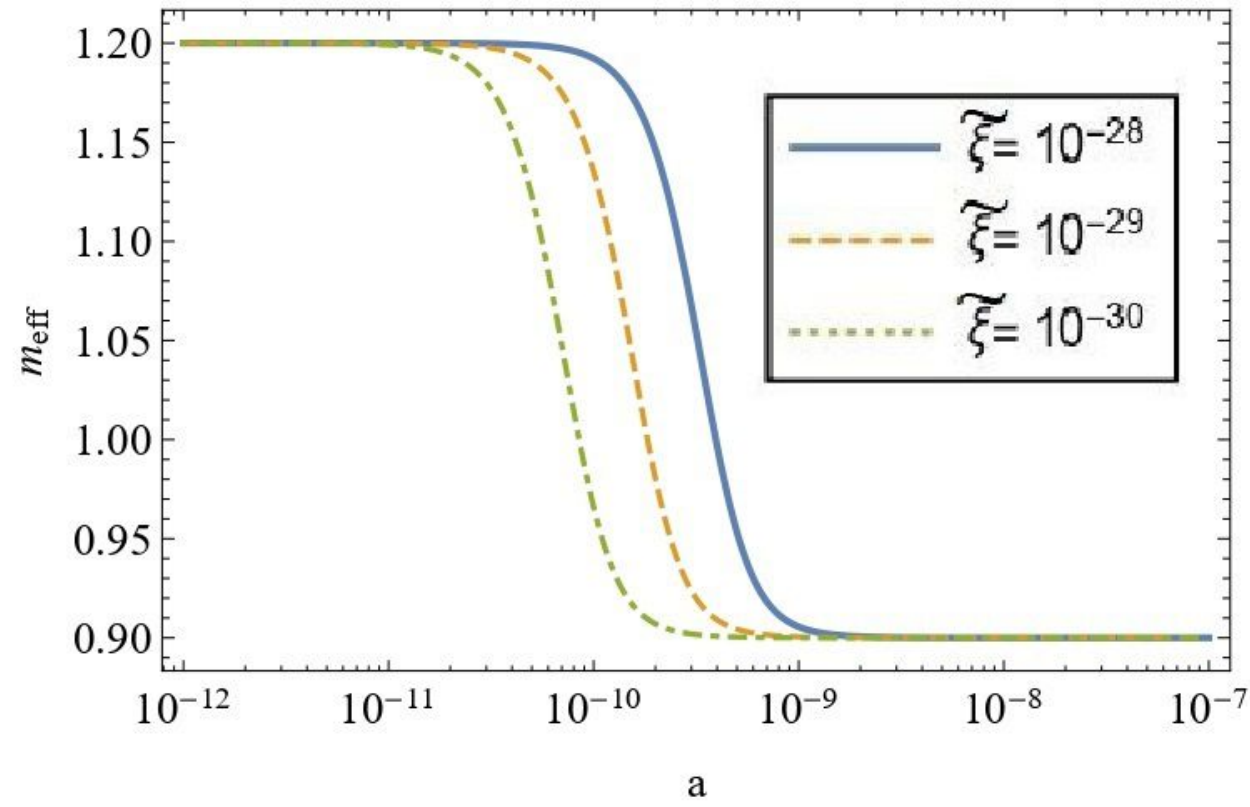}
\caption{{\it{The evolution of the normalized effective Newton's 
constant (upper graph), and of the normalized curvature-dependent fermion mass (lower 
graph), for various values of $\tilde{\xi}$, for 
 $ \rho_{f0} = 0.3$ and $ 
\Lambda= 0.7$, in units where $\kappa^2 = 1$.}
}} 
\label{fig3}
 \end{figure}

We close this section by examining the behavior of the effective Newton's 
constant $G_\text{eff}$ from (\ref{Geff}),  as well as the effective 
(curvature-dependent) fermion mass $m_{\text{eff}}$ from   (\ref{meffrel}), which using 
the Friedmann equation  (\ref{Friedmann}) becomes
\begin{equation}
\frac{m_{\text{eff}}}{\Lambda} = \frac{3 \tilde{\xi} \left[12 a^6+a^3 (14 
\tilde{\xi} +3 s)+8 \tilde{\xi}  s\right]}{2 \left(3 a^3+2 \tilde{\xi} \right)^2}+\frac{ 
m_0s }{ \rho_{f0}}.
\end{equation} 
In Fig.~\ref{fig3} we depict their normalized evolution, for various values of 
$\tilde{\xi}$. As we 
observe, at early times the curvature-dependent fermion mass is small, which makes the 
effective Newton's 
constant small too, and this effective ``weakening'' of gravity  is the reason for the 
inflation 
realization.
 In particular, the effective Newtonian constant starts from $0$ and results at $G_N$ 
(we mention that we can choose  $\tilde{\xi}$ in order to obtain a small 
post-inflationary variation of $G_\text{eff}$, in order to be in agreement with the Big 
Bang Nucleosynthesis (BBN) constraints \cite{Nesseris:2017vor,Kazantzidis:2018rnb}), 
while 
$m_{\text{eff}}$ 
starts from $\Lambda \left(3s+  \frac{ 
m_0s }{ \rho_{f0}} \right)$ and 
results to $2\Lambda   \tilde{\xi}$, in agreement with the fact that  fermionic matter is 
cold.

Nevertheless, apart from the above BBN bound on $G_\text{eff}$ itself, we 
should 
check the full BBN constraints on the scenario, i.e. focusing on the ratio of the standard
radiation to the above dark component. Although a complete analysis lies beyond the scope 
of the present work, we can  explicitly check the compatibility of fermionic inflation   
within the standard BBN using the average bound on the possible variation of the BBN 
speed-up factor \cite{Uzan:2010pm,Sola:2016jky}. 
First of all, since the model at hand includes only the fermionic sector, as we 
mentioned above if 
we want to additionally describe  the radiation epoch (and hence the BBN phase) we need 
to include a reheating mechanism that could produce the radiation sector. For the purpose 
of the BBN analysis we assume   that such a mechanism took place and thus 
at post inflationary universe  the   Friedmann equation  (\ref{Friedmann}) extends to 
\begin{equation}\label{Friedmannb}
\frac{3H^2}{\kappa^2}+\frac{2\tilde{\xi}H^2}{\kappa^2a^3}
= \frac{\rho_{f0}}{a^3}+\Lambda + \frac{\rho_{r0}}{a^4},
\end{equation}
with $\rho_{r0}$ a parameter. 
Now, the BBN 
speed-up factor
 is defined as the ratio of the difference of the
expansion rate predicted in a given model versus that of $\Lambda$CDM with standard 
radiation at the BBN epoch, i.e. at $z_{\rm BBN} \sim 10^9$ or at $a_{\rm BBN} 
=\frac{1}{z+1} \sim 10^{-9}$ (setting the current value $a_0=1$), namely 
$\Delta H^2/H_{\Lambda\mathrm{CDM}}^2 $ 
\cite{Uzan:2010pm,Sola:2016jky}. 
Thus, in our 
case, using  (\ref{Friedmannb}) (which for $\tilde{\xi}=0$ coincides with $\Lambda$CDM) 
we 
obtain
\begin{equation}
 \frac{\Delta H^2}{H_{\Lambda\mathrm{CDM}}^2} = 
\frac{2\tilde{\xi}}{2\tilde{\xi}+a_\text{BBN}^3} < 10\%.
\end{equation}
The above constraint yields the limit: $\tilde{\xi} < a_\text{BBN}^3/8 \sim 10^{-28}$.  
Hence, choosing  $\tilde{\xi}$ inside this bound, the  BBN requirements are safely 
satisfied in the context of fermionic inflation (indeed in all the above figures 
$\tilde{\xi}$ has been chosen accordingly).

\section{Perturbations}
\label{Perturbations}

In this section we investigate the perturbations of the above background evolution, and 
in 
particular we focus on the inflationary observables such as the   scalar spectral index 
$n_s$ and the tensor-to-scalar ratio $r$. As usual,  we introduce the Hubble 
slow-roll parameters $\epsilon = -\dot{H}/H^2$, $\eta   = 
-\ddot{H}/(2H \dot{H})$
\cite{Martin:2013tda,Bamba:2016wjm}, which in our case using the Friedmann equation  
(\ref{Friedmann}) read
\begin{eqnarray}
&&\epsilon \left(a \right) = \frac{3 s}{2 
\left(a^3+s\right)}-\frac{3\tilde{\xi} }{3 a^3+2\tilde{\xi} }, 
\\
&&
\eta  \left(a \right)=  \frac{3}{2}-\frac{6 \tilde\xi }{3 
a^3+2 \tilde\xi }.
\end{eqnarray}
Inflation ends  when $\epsilon (a_f) = 1$, and as mentioned above using the ``effective'' 
potential one can see that $a_f = a_B$, with $a_B$ the scale factor value 
at the   minimal point $B$ of Fig.~\ref{fig1}. Thus, if the initial scale factor 
is $a_i$ then the number of $e$-folding is
\begin{equation}
N = \log{\left(\frac{a_f}{a_i}\right)}.
\end{equation}

Using the slow-roll parameters, one can calculate the values of the scalar 
spectral index and the tensor-to-scalar ratio respectively as  
\cite{Nojiri:2019kkp,Dalianis:2018frf}
\begin{eqnarray}
&&
r = 16 \epsilon = \frac{24 s}{a_i^3+s}-\frac{48 \tilde{\xi}  }{3 a_i^3+2 \tilde{\xi}  }
\label{rrel}
\\
&&
n_s = 1- 6\epsilon + 2 \eta = 4 -\frac{9 s}{a_i^3+s}+\frac{6 \tilde{\xi} }{3 
a_i^3+2\tilde{\xi} }.
\label{nsrel}
\end{eqnarray}
As we can see, the first slow-roll condition $0<\epsilon \ll 1$ leads to 
\begin{equation}
0<a_i^3\leq \frac{4 \tilde{\xi}  }{3},\quad \frac{2}{3}\tilde{\xi}<s, \quad 0 \ll s,
\end{equation}
while the  second slow-roll condition  $0 < \eta \ll 1$ yields
\begin{equation}
\frac{2\tilde{\xi}  }{3} \leq a_i^3 \ll \frac{10 \tilde{\xi}  }{3}.
\end{equation}
Hence, combining them we obtain the requirement
\begin{equation}
\frac{2}{3} \tilde{\xi}  \lesssim a_i^3 < \frac{4}{3} \tilde{\xi} ,\quad 
\frac{2}{3}\tilde{\xi}  \lesssim s.
\label{con1}
\end{equation}
Finally, from (\ref{rrel}),(\ref{nsrel}) we can obtain 
\begin{equation}
r=\frac{8(4-n_s)}{3}-\frac{32 \tilde{\xi} }{3 
a_i^3+2\tilde{\xi} }.
\end{equation}
As one can see, by suitably
choosing the values of  $ \tilde{\xi}$ and $s$ we obtain $r$ and $n_s$   well 
inside the Planck observed values \cite{Akrami:2018odb}. For instance, taking according 
to (\ref{con1}) that $a_i^3\approx \frac{2}{3} \tilde{\xi}$ we obtain that 
$3r\approx8(1-n_s)$, which gives $n_s=0.97$ and $r=0.08$. This is one of the main results 
of the present work and reveals the capabilities of the scenario at hand, which can give 
a solution in agreement with observations, both at background and perturbation levels.

\section{Conclusions}
\label{Conclusions}
In this work we constructed a model of inflation realization, driven by fermions with 
curvature-dependent mass. We started from the  Covariant Canonical Gauge Theory of 
gravity 
 (CCGG) formulation, which imposes the requirement of invariance with respect to local 
transformations in curved spacetime through the insertions of gauge fields 
\cite{Struckmeier:2017vkf}. Incorporation of Dirac fermions in such a framework 
implies that the fermions acquire a curvature-dependent mass. The effect of such 
a correction is expected to be significant in regimes where the Ricci curvature is large, 
such is the case in the early universe. In particular, the dark spinor dust is indeed 
repulsive due to the $m_\text{eff}$ term, as long as we are close to 
a singularity, either close to black holes or close to the   Big Bang.   

We first investigated the scenario at the background level. As we showed, we obtained the 
inflation realization, with an initial de Sitter phase followed by a successful exit. 
Furthermore, we acquired the subsequent thermal history, with an effective matter era, 
followed finally by a dark-energy epoch. This behavior is qualitatively expected, since 
as we showed the curvature-dependent fermion mass induces a smaller effective Newton's 
constant at early times,  and this effective ``weakening'' of gravity  is the reason for 
the inflation realization.

We proceeded to the investigation of the scenario at the perturbation level, focusing on 
observables such as the scalar spectral index and the tensor-to-scalar ratio.
We extracted analytical expressions for their values, and we showed that by suitably 
choosing the coupling parameter, we can obtain values in agreement with Planck 
observations.
The efficiency of inflation from fermions with curvature-dependent mass at both the 
background and perturbation level reveals the capabilities of the scenario and makes it a 
good candidate for the description of nature.

One could try to investigate extensions of the above basic scenario. In particular, in 
this work we treated the spinor fields classically, assuming them to lie at the minimal 
state, namely being cold. Nevertheless, incorporation of high-temperature effects, as for 
example in the case of warm inflation 
\cite{Berera:1995ie,Kamali:2017zgg}, could lead to 
the appearance of an additional friction term, and this could further improve the values 
of the  spectral index and the tensor-to-scalar ratio.
Another possible generalization is to consider a quadratic curvature dependence, alongside 
the linear one, in the fermion mass, or even more general functions of the form 
$m_{\text{eff}} = f \left( m_0,R \right)$.  

We close this work by mentioning that in scalar field models of 
inflation the transition into the radiation and matter fields is realized  by introducing 
additional fields, such as the curavton field.
The exact way with which the inflationary fermions decay into other particles is an 
important question that should be investigated in a future project.

\section*{Acknowledgments}
D. B. thank the FIAS for generous support. D.V. thanks the Carl-Wilhelm Fueck Stiftung for 
generous support through the Walter 
Greiner Gesellschaft (WGG) zur Foerderung der physikalischen Grundlagenforschung 
Frankfurt. D.B.,  E.I.G. and E.N.S are grateful for the support of COST Action 
CA15117 ``Cosmology and Astrophysics Network for Theoretical Advances and Training 
Action'' (CANTATA) of the European Cooperation in Science and Technology (COST).
H.S. thanks the WGG and Goethe University for support through the Judah Moshe Eisenberg 
Laureatus endowed professorship and the BMBF (German Federal Ministry of Education and 
Research).

\end{document}